# Band Alignment in Black Phosphorus/Transition Metal Dichalcogenide Heterolayers: Impact of Charge Redistribution, Electric Field, Strain and Layer Engineering

*Nupur Navlakha[*], Priyamvada Jadaun, Leonard F. Register and Sanjay K. Banerjee*

Microelectronics Research Center, Department of Electrical and Computer Engineering,

The University of Texas at Austin, Austin, TX-78758, United States

KEYWORDS: Heterostructure, Band alignment, Electron Affinity, Biaxial Strain, Black Phosphorus, Molybdenum disulphide, Charge Redistribution, 2D Monolayers, Dielectric, TMD

**Abstract**: In this work, the energy band alignments of heterostructures of 2D materials are studied, where these are crucial for various device applications. Using density functional theory (DFT), we consider heterostructures of Black Phosphorus (BP) with transition metal dichalcogenides ($MX_2$), where M = Molybdenum (Mo), Tungsten (W), Hafnium (Hf) and X = Sulphide (S), Selenide (Se), and, specifically, the effects of charge redistribution and associated electrostatic fields on the band alignments beyond the electron affinity rule, as well as band tunability via applied layer-normal electric fields, applied strain, and layer engineering in $BP/MoS_2$. BP is a material with high mobility, mechanical flexibility, and is also sensitive to the number of BP layers. Absent such tuning, calculations for BP combined with the more electronegative materials result in a staggered (Type II) alignment for $MoS_2$, and a broken gap (Type III) alignment for $HfSe_2$, and $HfS_2$. Calculation for BP with less electronegative materials, $WSe_2$, $MoSe_2$ and $WS_2$ materials result in straddling (Type I) alignment, with a direct gap for $WSe_2$ and $MoSe_2$, and an indirect gap for $WS_2$. The amount of charge redistribution between layers and associated variations from the electron affinity rule increase going from Type I to Type II to Type III, where the band alignment becomes significantly pinned in the latter case by the creation of mobile charge carriers. With such tuning, these band alignments can then be altered quantitatively and qualitatively.

**INTRODUCTION**

Stacking of individual layers of two-dimensional (2D) materials has been introduced to achieve scalability and band tunability in various devices [1-3]. Monolayers of 2D materials couple to form a layered structure via relatively weak van der Waals (vdW) bonds. Such layered structures have demonstrated controllability of band alignments through strain, number of layers, chemical doping, alloying, and externally applied fields [1-7]. They have gained attention due to their ultra-thin bodies that show excellent electrostatic control, high mechanical flexibility, and absence of dangling bonds at the surface, which reduces interface traps and defects [1].

The type of application often depends on the type of band alignment. Heterostructure stacking can result in Type I (straddling), II (staggered) and III (broken) band alignments [2,3]. In vdW materials, a Type I alignment, where the conduction band minimum (CBM) and the valence band maximum (VBM) of the stack occur in the same/narrower bandgap material [3], produces confinement of electron and holes in the same region and thereby enhances radiative recombination and, thus, can be utilized for light emitting applications [3]. A Type II alignment, where the CBM and VBM are in different materials, is desirable for photovoltaics and photodetectors [3,8]. A Type III alignment, where the CBM of one material overlaps the VBM of the other material, may be utilized for tunnel diodes [3,9].

Black Phosphorus (BP) has potential applications in various optoelectronic devices due to its direct band gap, high mobility, flexibility, and tunability with number of layers, strain tolerance, and anisotropic physical properties [10-16]. Transition metal dichalcogenides (TMDs) have a sizable band gap unlike graphene and have demonstrated potential applications in optoelectronic and nanoelectronic devices [1-5,9,17-18]. Among the TMD family, molybdenum disulphide ($MoS_2$) is the most widely studied, and it exhibits many desirable material properties well suited for transistor applications, including relatively good carrier mobility [17,18]. BP, MoS$_2$, and their heterostructures have shown potential in a myriad of applications, such as non-volatile memory [19], photodetectors [8], rectifier diodes [20], and field-effect transistors [18, 21-22]. A vertically stacked *pn* junction formed with BP and MoS$_2$, hafnium disulfide (HfS$_2$), or hafnium diselenide (HfSe$_2$) may be usable for electron-hole bilayer (EHB) tunnel field effect transistors (TFETs), which have been difficult to realize using silicon (Si). [23]. Stacked BP and tungsten disulfide (WS$_2$), and stacked BP and molybdenum diselenide (MoSe$_2$) also can be tuned to operate as a *pn* junction, while stacked BP and tungsten diselenide (WSe$_2$) is well suited to function as a *p/p* or an *n/n* heterojunction [24]. Furthermore, a surface-normal electric field can provide electrostatic doping of these 2D materials [25] while also allowing for reconfigurable devices [26].

Improving and expanding the applicability of these heterostructures requires an in-depth analysis of interlayer interactions and the factors that influences the interfaces includes material and layer engineering, strain, and applied normal fields. In this work, using density functional theory (DFT), we study heterostructures of BP with the MoS$_2$, MoSe$_2$, WS$_2$, WSe$_2$, HfS$_2$, and HfSe$_2$ and, more specifically, the effects of charge redistribution and associated electrostatic fields on the band alignments beyond the electron affinity rule (Anderson's Rule) [27].

**COMPUTATIONAL METHOD**

Our calculations are performed using first-principles Density Functional Theory (DFT) with the projector-augmented wave (PAW) method as implemented using the Vienna ab initio Simulation Package (VASP) [28]. The exchange-correlation interaction is included within the generalized gradient approximation (GGA) developed by Perdew-Burke-Ernzerhof (PBE) [29]. The optimized lattice parameters of monolayers of BP [10,13,14] and TMDs are consistent with published literature [3,4] and are provided in Table I of supplementary information. Creating a 2D vdW heterojunction requires stacking at least two materials. The challenge for simulation is the formation of supercells with both limited lattice strain and computationally tractable numbers of atoms. We combined 1×4√3 supercells of $MoS_2$, $MoSe_2$, $WS_2$, $WeS_2$ with a 1×5 supercell of BP to create a heterostructure of 44 atoms (20 P atom, 8 M atom and 16 X atom), while 5×√3 supercells of $HfS_2$ and $HfSe_2$ are combined with a 4×2 BP to create a heterostructure of 62 atoms (32 P, 10 Hf, 20 S/Se) (Figure 1a). That said, the strain required to realize practical unit cell sizes can quantitatively affect the band structures of the considered materials. Therefore, while we necessarily provide specific values for various quantities below, our focus is on qualitative effects.

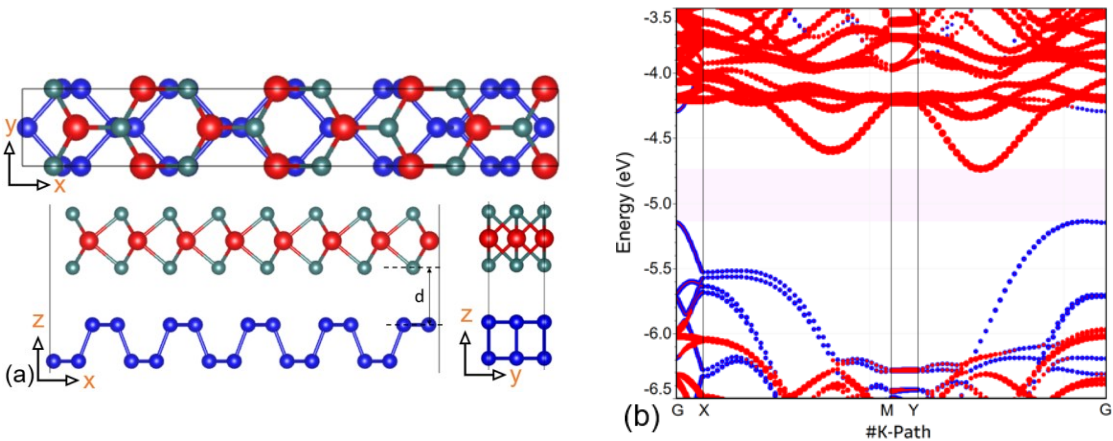

**Figure 1.** (a) The crystal structure and (b) projected band structure of BP/$MoS_2$ heterostructure. In (a) the *z* direction is taken as normal to the plane of the $MoS_2$ and BP layers, and BP, Mo, and S atoms are represented in blue, red, and green, respectively. In (b) blue and red dots represent the projection of each energy band onto the BP and $MoS_2$ layers, respectively, where the size of dots corresponds to the weight of the projections (with the red dots overlaying the blue ones where they overlap in position). All energies are referenced to vacuum level.

Van der Waals interactions, modeled using the OptB88 functional method [30], are used to calculate the interfacial distance between the X atom-center to the P atom-center (denoted as "*d*" in Figure 1(a)) and binding energy ($E_b$) between the layers. $E_b$ is calculated as $E_b = E_{BP/TMD} - E_{BP} - E_{TMD}$, where $E_{BP/TMD}$ is the total energy of BP/TMD heterostructure and $E_{BP}$ and $E_{TMD}$ are the total energies for isolated monolayers of BP and TMDs, respectively. The thickness of the remaining vacuum region outside the heterostructure in the simulation is greater than 15 Å. The structure was fully relaxed with a force tolerance of 0.01 eV/Å before calculation of the electronic properties of the heterostructure [31]. The energy cutoff was set to 400 eV, and Brillouin zone was sampled using Monkhorst-Pack grids of 5 × 21 × 1 for bilayers composed of BP with $MoS_2$, $MoSe_2$, $WS_2$ or $WeS_2$, and 6 × 11 × 1 for bilayers of BP with $HfS_2$ or $HfSe_2$. The break criterion for the electronic self-consistent loop was set to $10^{-5}$ eV. The resulting heterostructure lattices and their binding energies are included in Table II of supplementary information.

**RESULTS AND DISCUSSION**

The band alignment between layers of the vdW stacks can be altered using different materials, different layer numbers, externally applied electric fields, and strain. Moreover, even with zero applied external field, charge redistribution between the layers can affect band alignment between layers, which includes band offsets and the interlayer band gap, and the electron affinity of the heterostructure, all important to device and material applications.

*A. Heterostructure Band Alignment (zero external field)*

Figure 1 shows the BP/$MoS_2$ heterostructure with its projected band structure. It is evident from the figure that the CBM of the heterostructure originates from $MoS_2$, while VBM of the heterostructure originates from BP, indicating a Type II alignment. The BP/$MoS_2$ heterostructure has an indirect band gap with the CBM and the VBM located between Y and Γ (the latter labeled "G") but at different **k**-points and a simulated band gap, $E_g = E_c - E_v$, of 0.407 eV. This value is comparable to those of previous works [21,31,32]. The band structure of the other heterostructures is shown in the supplementary data (S1), while the band alignments of the stacked system is included in Table 1 and shown in Figure 2(a).

**TABLE 1.** Band parameters for strained monolayers of BP and TMDs. $\Delta E_c = E_{c\_BP} - E_{c\_TMD}$, $\Delta E_v = E_{v\_BP} - E_{v\_TMD}$ are the band offsets for CB energies and VB energies, respectively. In the simulated heterostructure, BP and the TMD layer-projected band edge energies are used for $E_c$ and $E_v$. $E_{g(Het)}$ is the bandgap obtained from simulation of the heterostructure as a whole, while $E_{g(EAR)}$ is the bandgap of heterostructure obtained from the simulation of the isolated (but identically strained) monolayers using the electron affinity rule (EAR). The text highlighted in grey indicates the material and associated band edge energy defining the CBM and VBM of the heterostructure. All energies are in eV and band-edge energies are reference to the vacuum level.

| (Units of eV) | | Electron Affinity Rule (based on isolated monolayers) | | | | | Simulated heterostructure (layer projected) | | | | | |
|---|---|---|---|---|---|---|---|---|---|---|---|---|
| Stack | Layer | $E_c$ | $E_v$ | $\Delta E_c$ | $\Delta E_v$ | $E_g$ | $E_c$ | $E_v$ | $\Delta E_c$ | $\Delta E_v$ | $E_g$ | $E_{g\,(Het)} - E_{g\,(EAR)}$ |
| MoS$_2$/BP | MoS$_2$ | -4.79 | -6.14 | 0.30 | 1.04 | 0.31 | -4.74 | -6.04 | 0.44 | 0.89 | 0.40 | 0.09 |
|  | BP | -4.49 | -5.1 |  |  |  | -4.30 | -5.15 |  |  |  |  |
| MoSe$_2$/BP | MoSe$_2$ | -4.16 | -5.60 | -0.17 | 0.36 | 0.92 | -4.17 | -5.61 | -0.08 | 0.36 | 0.99 | 0.07 |
|  | BP | -4.33 | -5.248 |  |  |  | -4.26 | -5.251 |  |  |  |  |
| WS$_2$/BP | WS$_2$ | -4.43 | -5.97 | -0.08 | 0.86 | 0.61 | -4.41 | -5.95 | -0.03 | 0.84 | 0.67 | 0.06 |
|  | BP | -4.5 | -5.111 |  |  |  | -4.44 | -5.114 |  |  |  |  |
| WSe$_2$/BP | WSe$_2$ | -3.84 | -5.38 | -0.49 | 0.14 | 0.91 | -3.87 | -5.41 | -0.40 | 0.17 | 0.98 | 0.07 |
|  | BP | -4.33 | -5.238 |  |  |  | -4.26 | -5.241 |  |  |  |  |
| HfS$_2$/BP | HfS$_2$ | -6.09 | -7.06 | 1.51 | 1.99 | -1.01 | -5.78 | -6.74 | 0.7 | 1.07 | -0.11 | 0.9 |
|  | BP | -4.59 | -5.08 |  |  |  | -5.09 | -5.67 |  |  |  |  |
| HfSe$_2$/BP | HfSe$_2$ | -5.70 | -6.57 | 1.26 | 1.36 | -0.49 | -5.49 | -6.29 | 0.82 | 0.81 | -0.01 | 0.48 |
|  | BP | -4.44 | -5.21 |  |  |  | -4.67 | -5.48 |  |  |  |  |

BP/WSe$_2$ and BP/MoSe$_2$ show Type I heterostructure band alignments with direct band gaps of 0.98 eV and 0.99 eV, respectively, located at Γ. BP/WS$_2$ also shows a Type I heterostructure but has an indirect band gap of 0.67 eV with the CBM at Γ and the VBM located at a **k**-point between Y and Γ. BP/HfSe$_2$ and BP/HfS$_2$ form a Type III heterostructure with an overlap of 0.01 and 0.11 eV, respectively between $E_c$ and $E_v$ with CBM and VBM located at different **k**-points between Y and Γ.

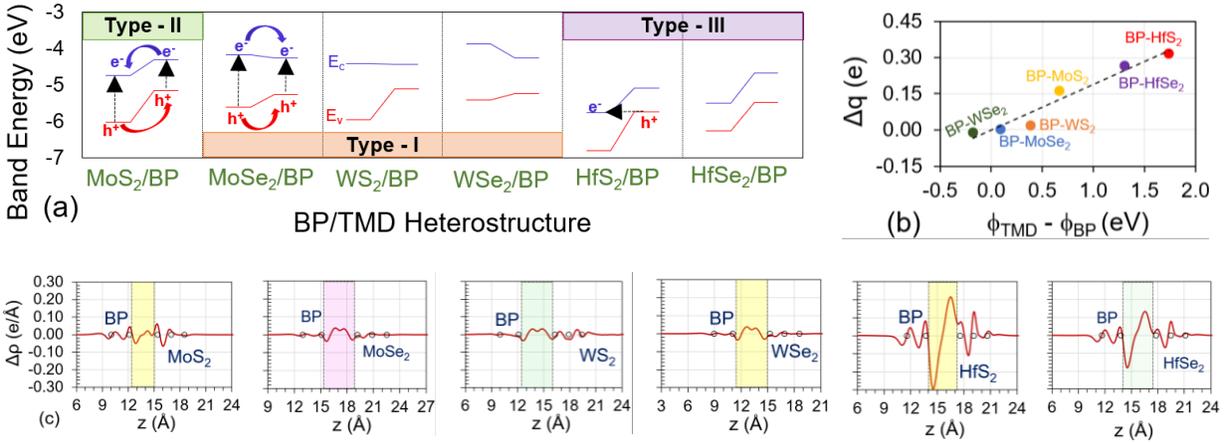

Figure 2. (a) DFT-calculated band alignment of BP/TMD heterostructures relative to the vacuum level, along with schematic illustration of the migration of photogenerated electrons and holes (such as in photo diodes) in Type I and II heterostructures, and of tunneling in Type III heterostructures. (b) Variation in the charge redistribution (in units of electron charge magnitude) between layers (Δq) as a function of the workfunction difference between the TMD monolayer ($\phi_{TMD}$) and the BP monolayer ($\phi_{BP}$). Positive Δq indicates electron redistribution from the BP to the TMD. (c) The *x-y* plane-averaged electron density difference along the *z* direction of the BP/TMD heterolayer. The shaded region in (c) represent the interlayer spacing (*d*).

For each heterostructure, Table 1 shows: the simulated valence and conduction band edge energies for electronically isolated (but still strained, consistent with the heterostructure) BP and TMD materials, and the corresponding band edge offsets and heterostructure band gaps to be expected from the electron affinity rule for reference; the same obtained from the calculated layer-projected apparent band-edge energies in the heterostructure; and the difference between these two bandgap calculations for each heterostructure. As seen, the type of heterostructure formed remains the same by either approach, but the predicted band edge offsets and bandgaps vary significantly, with energy differences varying from a few tens of meV to almost an eV for $HfS_2$/BP.

These differences between the expectations of the electron affinity rule and the as-calculated heterostructure band structures can be attributed significantly to charge redistribution—but not free carrier transfer in the still gapped Type I and II undoped systems at 0 K—within and between the layers resulting from heterostructure formation, as shown in Figure 2(b) and (c), although distortion of the orbitals may make a quantitative contribution as well, and the effects cannot be entirely decoupled. The work function for each isolated but strained monolayer is evaluated as in [5], and the charge transfer between layers within the heterostructure of Figure 2(b) are calculated through Bader charge analysis [33]. Figure 2(c) is a plot of charge density difference along z, $\Delta \rho\ (z)$, defined as,

$$\Delta \rho\ (z) \equiv \int \rho_S\ (x,y,z) dx dy - \int \rho_{BP}\ (x,y,z) dx dy - \int \rho_{TMD}\ (x,y,z) dx dy,$$

where $\rho_s$ is the density of the heterostructure stack, and $\rho_{BP}$ and $\rho_{TMD}$ are the densities of the isolated BP and TMD layers, respectively. The amount of redistribution between layers increases going from Type I to Type II to Type III. The results of Figure 2(b) show a near linear variation between charge transfer between the coupled TMD and BP layers and the difference in the work functions of the isolated layers.

Note, however, that unlike the case for an interface between two semi-infinite pieces three-dimensional materials, it cannot be assumed that band alignment between these 2D materials is independent of their environment of surrounding materials. For example, for a test heterostructure of $MoS_2$-BP-$MoS_2$, the heterostructure band gap is increased by 55 meV in these simulations (qualitatively opposite what would be expected from any state-splitting due to coupling between the electronic states of the MoS$_2$ layers through the BP), still further from the electron affinity rule. Put another way, changing the environment on one side of the BP from vacuum to MoS$_2$ has a significant impact on the band alignment on the other side. However, when we considered a test heterostructure of BP-MoS$_2$-BP, the calculated band alignments were almost same. Similar conclusions can be derived by considering a free-standing BP, BP-MoS$_2$, MoS$_2$-BP-MoS$_2$ and a free standing MoS$_2$, MoS$_2$-BP and BP-MoS$_2$-BP, where the projected band gap of BP changes in each case by approximately 200 meV with change in environment, while for MoS$_2$, the projected band gap is altered by 20 meV. To understand this relative insensitivity in the latter case, note that the charge redistribution of Fig. 2(c) occurs primarily between and about the most proximate atoms of the two intrinsic material layers, which are the chalcogenide atoms in the case of MoS$_2$, as can be seen if Fig. 2(b), while the near-band-edge band structure of TMDs is dominated by the central transition metal atoms [7]. These findings are consistent with the results of prior works [6] and suggest that TMDs may be unique in offering such strong isolation between the opposite sides of 2D materials.

### B. Impact of Applied Electric field

The impact of an applied electric field oriented normal to the layer stack, $E$, on the electrostatic potential and the band gap of the heterostructure, $E_g$, is now addressed. A positive (negative) applied field—with the field direction defined as positive from BP to MoS$_2$—produces a potential variation, as seen in Figure 3(a), and a net electron shift to (from) the BP from (to) MoS$_2$. The applied field is significantly screened within the bilayer, and even more so within the individual MoS$_2$ and the BP material layers. The effective static/low frequency dielectric constant for the bilayer (defined here considering the thickness of the bilayer as that of two material layers as well as the vdW gap between the layers and one-half that vdW gap on each side of the bilayer system) is approximately 4.5 and varies slightly with applied normal electric field (Figure 3(b)). We note that when we allow the atomic structure to relax with the applied electric field, we didn't find a significant change in the dielectric constant within the margin of error of the simulations. Despite this dielectric screening, the vdW stack bandgap can be adjusted quantitively and qualitatively among bandgap types with reasonable applied fields, as shown in Figure 3(c).

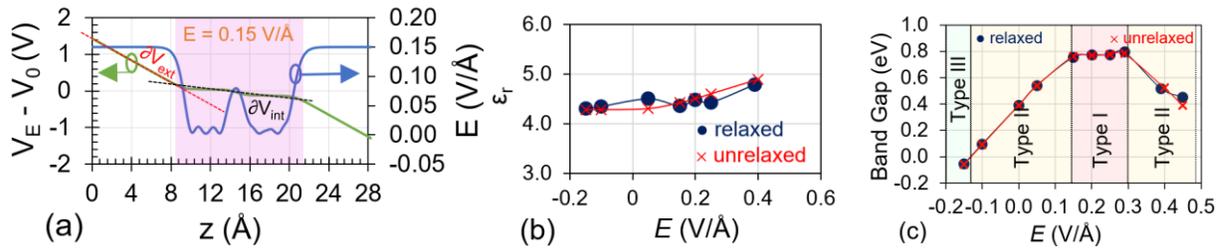

**Figure 3** (a) $x$-$y$ plane-averaged potential and electric field as a function of position $z$ for an applied electric field ($E$) of 0.15 V/Å normal to the stack, where the pink shaded region corresponds to the regions of the MoS$_2$ layer as defined in the text, (b) Relative dielectric constant ($\varepsilon_r$) of the heterostructure as a function of applied electric field obtained from $\varepsilon_r = E_{ext}/E_{int}$, where $E_{ext} = -\partial V_{ext}/\partial z$ and $E_{int} = -\partial V_{int}/\partial z$ are the average values of the MoS$_2$ external and internal fields, respectively, as illustrated in (a) via the dashed lines superimposed on the $x$-$y$ plane-averaged potential. (c) The band gap and type of the BP/MoS$_2$ heterostructure as a function of a normal electric field. The heterostructure can be tuned from Type III at negative fields (approximately $\leq -0.13$ V/Å) to Type I at positive fields (approximately $\geq 0.15$ V/Å). In (b) and (c), "relaxed" and "unrelaxed" refer to allowing the atomic structure to relax under the applied electric field or not, respectively. (Note that for the results of this figure (only), slightly altered DFT mixing parameters were used to achieve converge under all considered applied fields.)

A positive electric field of approximately 0.15 V/Å converts the heterostructure to Type I with the bandgap defined by BP, while a negative applied electric field of approximately $\leq -0.13$

V/Å, leads to a metallic Type III structure, with the valence band of BP slightly overlapping the conduction band of MoS$_2$. This bandgap tunability within, the Type II range through application of an external field offers the possibility of rectifier diodes with tunable barrier height [1], reconfigurable FETs [24], tunnel FETs [22], and electro-optical modulators [35].

## C. Strain Engineering

Strain-dependent modulation of transport properties have been studied and utilized for improving device performance in terms of mobility, tunability and control of magnetic properties [7,11]. Here, tensile (positive) and compressive (negative) strain, $(a-a_0)/a_0$, are considered. (To achieve compressive strain in practice, the BP-MoS$_2$ layer-normal ($z$) displacement would have to be constrained within a larger structure). Studies have shown that a semiconductor-metal transition in MoS$_2$ is predicted for a biaxial tensile strain of about 8% and compressive strain of about 15% [36], while for BP this transition is observed for compressive strain of about 9% and tensile strain of ~20% [37]. (Experimentally, MoS$_2$ can sustain uniaxial tensile strain greater than 11% [17], and BP has demonstrated capability to withstand tensile strain up to 30% in the zigzag direction and 27% in the armchair direction [11,38]).

To understand the impact of biaxial strain on BP/MoS$_2$ stack and the origin of the energy bands, we projected the Bloch states onto the atomic orbital basis of the constituent atoms as a function of energy, as shown in Figure 4. Considering the apparent bandgaps in Figure 5 as a function of strain and atom-projected Density of States (DOS) in MoS$_2$ and BP, the visually apparent VBM in MoS$_2$ is dominated by the $d_{z2}$ orbital of Mo and, to a lesser degree, the $p_z$ orbital of S [7]. The apparent CBM within MoS$_2$ is dominated by the $d_{z2}$ orbitals of Mo. For BP, the apparent VBM is dominated by the $p_z$ orbital. The apparent CBM of BP also is contributed primarily by the $p_z$ orbital, with a significant contribution from $p_x$, $p_y$ and s orbital as well.

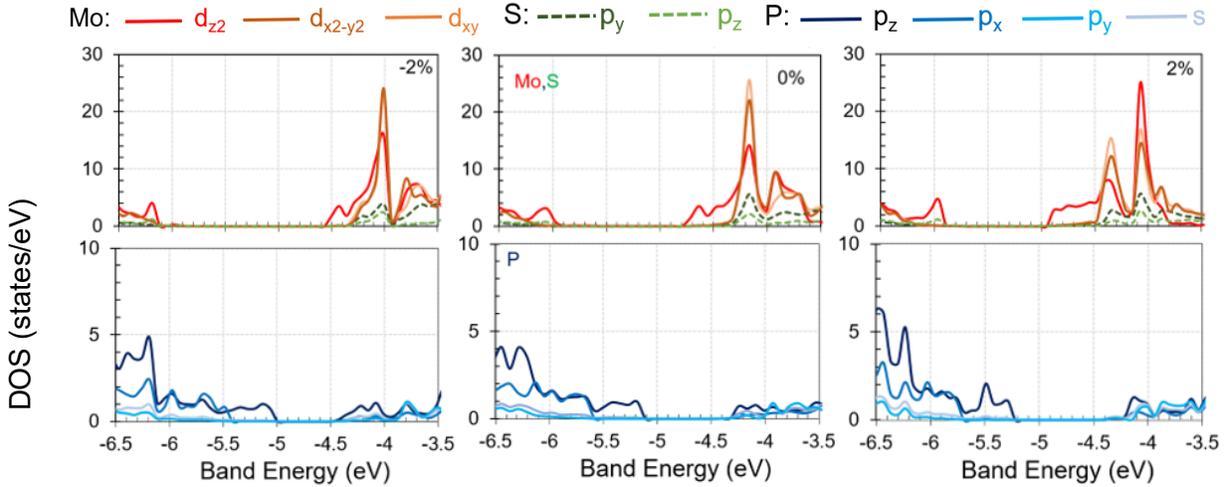

**Figure 4** Projected Density of States (DOS) for orbitals in BP/MoS$_2$ with biaxial strain of −2%, 0%, 2% for: $d_{z2}$, $d_{xy}$, $d_{x2-y2}$ of Mo; $p$ and s-orbital of P (blue), and $p_y$ and $p_z$ of S; where positive strain values correspond to tensile strain and negative values to compressive strain.

Figure 5(a) shows the change in band offset with applied strain, and the projected band gap of BP and MoS$_2$ is shown in Figure 5(b). As mentioned previously, the bond length increases with biaxial strain, and the MoS$_2$ bandgap decreases, consistent with reduced orbital overlap among the atoms. However, the band gap of BP increases with biaxial tensile strain due to decrease in distance between two sub-layers (height of a layer in $z$-direction) of phosphorus atoms within the puckered BP structure, as has been demonstrated experimentally [11] and explained theoretically [11,16] for isolated BP layer. The bandgap of MoS$_2$ varies more with tensile strain than compressive, and the opposite is true for BP. For the stacked system, Figure 5(b) shows that as strain varies from −5% to 5%, the band gap of BP/MoS$_2$ first increases, following the atom projected bandgap of BP in a region of strain where the BP/MoS$_2$ stack is Type I. Then, in the vicinity of and above −2% strain the BP/MoS$_2$ bandgap begins to decrease again as the stack becomes Type II with the valence band originating from the BP and the conduction band originating from the MoS$_2$.

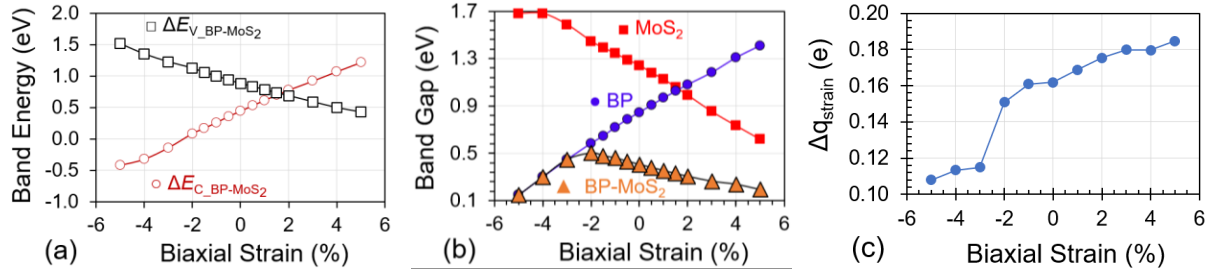

**Figure 5** Impact of biaxial strain on (a) CB and VB band offsets (as defined in Fig. 5(a)), and (b) band gap of BP, MoS$_2$, and BP/MoS$_2$ in the heterostructure, where positive strain values correspond to tensile strain and negative values to compressive strain. (c) Postive charge redistribution from MoS$_2$ to BP (corresponding to electron redistribution from BP to MoS$_2$) with the applied strain in units of the magnitude of the electron charge per supercell.

The charge distribution between the monolayers (Figure 5(c)) shows the electron loss from BP to MoS$_2$ increases by 0.03 electrons per supercell, from 0.16 to 0.19 for 5% (tensile) biaxial strain relative to that of the unstrained BP/MoS$_2$ stack, while the electron loss decreases by 0.05, from 0.16 to 0.11 for −5% (compressive) strain. Moreover, with tensile strain, the **k**-space position of the VBM of the BP/MoS$_2$ stack, again originating from the BP, shifts towards the Γ-point, while for compressive strain it shifts towards the Y-point. As for the CBM, while the BP/MoS$_2$ stack remains Type II with strain, the CBM originates from MoS$_2$, and the **k**-space position located along Γ to Y is essentially independent of strain. However, when the BP/MoS$_2$ stack becomes Type I with compressive strain and the CBM also originates from the BP, the **k**-space position of the CBM shifts to, and remains at, the Γ-point.

*D. Layer Engineering*

Use of multilayer BP has proven beneficial due to its direct band gap, high mobility, anisotropic electronic properties, and higher sensitivity as compared to other 2D materials including monolayer BP [6,10,12,15]. Few-layered BP can perform better than single-layer BP due to higher carrier density, lighter carrier effective mass and weaker scattering [6,12]. Type III heterostructures can be obtained by combining layer engineering with an externally applied field, in-plane strain, vertical stress, or plane-normal compression [39]. Multi-layer BP with $MoS_2$ could be useful for low voltage tunnel-FETs. Multi-layer BP with $WSe_2/MoS_2$ have shown applicability for MIR light-emission applications [40]. The change in the workfunction of BP with the stacking of BP layers possibly could be utilized for tuning contact resistance [12,41].

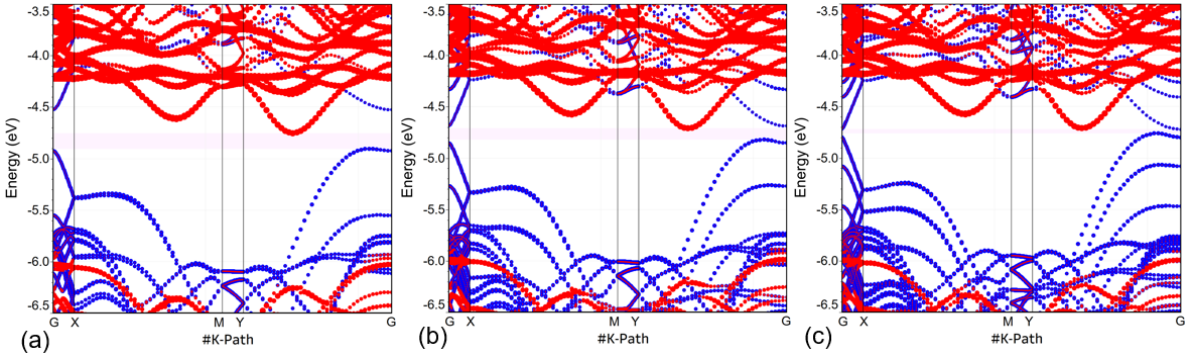

**Figure 6** Projected band energy of (a) 2L-BP/$MoS_2$, (b) 3L- BP/ $MoS_2$, and (c) 4L-BP/ $MoS_2$ heterostructure.

The variation in the band energy and alignment of the heterostructure with the number of BP layers varying from 2 to 4 is shown in Figure 6(a)-(c), respectively. Interlayer coupling between different layers of monolayer BP leads to band splitting, which shifts the CBM and the VBM to lower and higher energies, respectively [6,15]. As a result, the band gap decreases with increased number of BP layers [12,15], varying from 0.4 eV in one layer (1L) BP/$MoS_2$ to only 0.03 eV in four-layer (4L) BP/$MoS_2$. Figure 6(d) shows the plane-averaged electron density difference along the z direction of the BP/MoS$_2$ heterolayer. Table 2. provides the projection of the CBM and VBM onto the individual layers, indicating that these states are substantially delocalized with respect to BP layers, and, furthermore, the CBM becomes decreasingly localized to the MoS$_2$ layer as the layer projected CBM of the BP layers approaches that of the MoS$_2$ layer, although the VBM become increasingly localized to the BP layer with increasing number of BP layers.

**Table 2**. Contribution of each layer to the CBM and VBM states for *n*-layer-BP/MoS$_2$. These results were obtained through projection of the wavefunctions onto the atomic orbitals within the individual layers. Layer L1 is that closest to the MoS$_2$ layer, L2 the next closest, and so forth. The values are in percentage.

|  | VBM | | | | | CBM | | | | |
|---|---|---|---|---|---|---|---|---|---|---|
| **Stack** | MoS$_2$ | L1 | L2 | L3 | L4 | MoS$_2$ | L1 | L2 | L3 | L4 |
| 1L-BP/ MoS$_2$ | 16.3 | 83.7 | - | - | - | 91.5 | 8.5 | - | - | - |
| 2L-BP/ MoS$_2$ | 5.8 | 48.2 | 46 | - | - | 79 | 9 | 12 | - | - |
| 3L-BP/ MoS$_2$ | 4.1 | 35.2 | 42.2 | 18.5 | - | 58 | 7 | 16.3 | 18.7 | - |
| 4L-BP/ MoS$_2$ | 2 | 18.2 | 32 | 28.1 | 19.7 | 52.4 | 4.8 | 12.4 | 14.5 | 15.9 |

**CONCLUSION**

This study focused on understanding the impact of charge redistribution, external and internal fields, strain, and material and layer engineering on the band alignment of heterostructure through DFT simulations. Charge density analysis demonstrates electron redistribution between BP and the TMD layers, from BP to the TMD except in the case of WSe$_2$ where the redistribution is in the opposite direction, under no applied bias due to workfunction difference between the materials. This redistribution substantially alters the band alignment quantitatively from the expectations of the electron affinity rule, although the alignment type remains the same. By either approach, a Type I alignment is obtained for BP/WSe$_2$, BP/MoSe$_2$, and BP/WS$_2$, Type II alignment is obtained for BP/MoS$_2$ and a Type III alignment for BP/HfS$_2$ and BP/HfSe$_2$ under our simulation conditions. Adding biaxial strain affects both the interlayer and intralayer charge distribution and coupling among orbitals due to variation in bond lengths and angles. Moreover, simulation of BP/MoS$_2$ heterostructures suggest that band alignment can be tuned among all types, I, II and III, via an applied layer-normal electric field, and that it can be varied widely with strain or by increasing the number of BP layers. Such tunability offers flexible application to electron-hole bilayer-TFETs, reconfigurable FETs, electro-optical modulators, rectifier diodes, contacts, photo conductors and light emitting devices.

AUTHOR INFORMATION

**Corresponding Author**

*Email: nupurnavlakha@utexas.edu

ACKNOWLEDGMENT

We would like to thank National Nanotechnology Coordinated Infrastructure (NNCI), which is supported by the National Science Foundation under the Grant ECCS-2025227, which is. We also acknowledge the Texas Advanced Computing Center (TACC) at The University of Texas at Austin for providing high performance computing resources for this work.

[19] Liu, Chang, et al. "Polarization-Resolved Broadband MoS$_2$/Black Phosphorus/MoS$_2$ Optoelectronic Memory with Ultralong Retention Time and Ultrahigh Switching Ratio." *Advanced Functional Materials* 31.23 (2021): 2100781.

[20] Deng, Yexin, et al. "Black phosphorus–monolayer MoS$_2$ van der Waals heterojunction p–n diode." *ACS nano* 8.8 (2014): 8292-8299.

[21] Huang, Le, et al. "Electric-field tunable band offsets in black phosphorus and MoS$_2$ van der Waals pn heterostructure." *The journal of physical chemistry letters* 6.13 (2015): 2483-2488.

[22] Wu, Peng, et al. "Complementary black phosphorus tunneling field-effect transistors." *ACS nano* 13.1 (2018): 377-385.

[23] Agarwal, Sapan, et al. "Engineering the electron–hole bilayer tunneling field-effect transistor." *IEEE Transactions on Electron Devices* 61.5 (2014): 1599-1606.

[24] Chen, Peng, et al. "Gate tunable WSe$_2$–BP van der Waals heterojunction devices." *Nanoscale* 8.6 (2016): 3254-3258.

[25] Lu, Anh Khoa Augustin, et al. "Toward an understanding of the electric field-induced electrostatic doping in van der Waals heterostructures: A first-principles study." *ACS applied materials & interfaces* 9.8 (2017): 7725-7734.

[26] Mikolajick, Thomas, et al. "The RFET- A reconfigurable nanowire transistor and its application to novel electronic circuits and systems." *Semiconductor Science and Technology* 32.4 (2017): 043001.

[27] Davies, J. H., *The Physics of Low-Dimensional Semiconductors*. UK: Cambridge University Press (1997).